\documentclass{emulateapj}

\shorttitle{A New Low Surface Brightness Galaxy}
\shortauthors{Liang et al.}

\begin{document}
   \title{SDSS J121811.0+465501.2:~ a new Low Surface Brightness Galaxy with low metallicity}
  
   \author {Y. C. Liang\altaffilmark{1}, J. Y. Hu\altaffilmark{1}, F. S. Liu\altaffilmark{1}, Z. T. Liu\altaffilmark{2}}

\altaffiltext{1}{National Astronomical Observatories, Chinese Academy of Sciences, 
       20A Datun Road, Chaoyang District, Beijing 100012, China; 
       emails: ycliang@bao.ac.cn, hjy@bao.ac.cn, lfs@bao.ac.cn}
\altaffiltext{2}{School of Electronics and Information Engineering, Beijing
    Jiaotong University, Beijing 100044, China; email: liuzht@bjtu.edu.cn}


\begin{abstract}
   We serendipitously find a new nearby Low Surface Brightness (LSB)
   galaxy from SDSS database. We estimate oxygen abundance 
   of its H~{\sc ii} region SDSS J121811.0+465501.2 from electron temperature,
   as well as for another H~{\sc ii} region, SDSS J135440.5+535309.6,
   located in irregular LSB galaxy UGC\,8837. 
   These two extragalactic H~{\sc ii} regions were classified
   as stars in the SDSS-DR4 database, and were found occasionally by us 
   in the automatic recognition and classification on stellar spectra.
   Their optical spectra show obvious emission lines, i.e.,
  strong [O~{\sc iii}]$\lambda$$\lambda$4959, 5007, 
  Balmer emission lines, 
  but very weak [N~{\sc ii}]$\lambda$$\lambda$6548,6583
  and [S~{\sc ii}]$\lambda$$\lambda$6317,6731,
  which could indicate that they are metal-poor star-forming regions.
  The derived oxygen abundances of the two objects are
   12+log(O/H)$\sim$ 7.88$\pm$0.30 and 7.70$\pm$0.30, 
   respectively.
   The host of the H~{\sc ii} region SDSS J121811.0+465501.2 
   is identified as a new nearly edge-on LSB disc galaxy (almost without bulge)
   with the $B$-band central surface brightness $\mu_0(B)$ as 23.68 mag arcsec$^{-2}$
   and inclination angle as $\sim$75 degree by using
   the GIM2D software to analyze its $g-$ and $r-$band images independently.
   It is a nearby dwarf galaxy with redshift $z\sim$0.00157, disk scale-length
   $\sim$0.40\,kpc and $B$-band absolute magnitude $M_B\sim -$13.51 mag.
   The very low oxygen abundances of these two objects 
   confirm the low metallicities of LSB galaxies.
\end{abstract} 
  \keywords{galaxies: abundances ---
          galaxies: dwarf ---
          galaxies: evolution ---
          galaxies: ISM ---
	  galaxies: irregular
          galaxies: spiral ---
          }
%

%
\section{Introduction}

The Sloan Digital Sky Survey (SDSS, York et al. 2000) 
has provided a large set of unique data
to study the properties of astrophysics objects, such as
galaxies, stars, quasars.
Many projects have been done or are ongoing on the basis of
this large dataset. For example, about galaxies, besides   
the large number of typical studies, such as 
calculating the galaxy luminosity function (Blanton et al. 2001),
studying the statistic properties of bright galaxies (Shimasaku et al. 2001),
the color separation of galaxy types (Strateva et al. 2001),
the mass-metallicity relations and star-formation history 
of star-forming galaxies (Kauffmann et al. 2003a,b; 
Brinchmann et al. 2004; Tremonti et al. 2004),
the metallicity calibrations of star-forming galaxies (Liang et al. 2006; Yin
et al. 2007), 
the color-magnitude relations of galaxies (Chang et al. 2006a,b) etc.,
especially, the SDSS database is also very efficient
to search for special interesting objects 
by taking advantage of the high-quality optical spectra 
and deep multi-bands photometric images 
(Wang et al. 2006; Kewley et al. 2007; Zhang et al. 2007). 

 We found two extragalactic H~{\sc ii} regions from 
the SDSS data release 4 (DR4) database,
which were classified as stars in the database. Indeed,
one of them, SDSS J121811.0+465501.2, is located in 
a galaxy which is identified as a new Low Surface Brightness (LSB) 
galaxy by this study (Sect.\ref{sec5}), 
and another one, 
SDSS J135440.5+535309.6, is located in the irregular LSB galaxy UGC\,8837.
The reasons they were classified as stars  
may be that their hosts are very faint with very low luminosities 
and low surface brightness, and they have very low redshifts. 
However, their SDSS optical spectra show obvious emission lines, i.e.,
strong [O~{\sc iii}]$\lambda$$\lambda$4959, 5007, 
Balmer emission-lines at low and high levels, 
but very weak [N~{\sc ii}]$\lambda$$\lambda$6548,6583
and [S~{\sc ii}]$\lambda$$\lambda$6317,6731,
which show that they could be metal-poor star-forming regions.
These two objects must be very interesting to check the metallicities of
LSB galaxies by taking advantage of the high quality spectral observations.

  LSB galaxies are galaxies  that emit much less light
 per area than normal galaxies. Because of their contrast with  the night sky they
 are hard to find, and hence their contribution to the general galaxy population 
 has long been underestimated. 
 They are usually defined as objects with a blue central surface brightness 
 $\mu_0(B)$ significantly fainter than the Freeman value of 21.65 mag arcsec$^{-2}$
 (Freeman 1970). This threshold value of  $\mu_0(B)$ 
 varies
 in the literature from $\geq$23.0 mag arcsec$^{-2}$ (Impey \& Bothun 1997) to 
   $\geq$22.0 mag arcsec$^{-2}$ (Impey, Burkholder \& Sprayberry 2001).
 McGaugh (1996) tried to quantify this as well.
 They described as an LSB any
 disc galaxy with $\mu_0(B)$ $>$ 22.75 mag arcsec$^{-2}$.
 Further quantitative distinctions were also suggested 
 to avoid the pollution of 
 excessive nomenclature, for example, 
 the intermediate surface brightness (ISB) with
 $22<$$\mu_0(B)$ $<$ 22.75 between LSB and high surface brightness (HSB),
 the very low surface brightness (VLSB) 
 with $24.5<$$\mu_0(B)$ $<$ 27 and so on.   
      The detection of LSB galaxies is difficult owing to their intrinsically 
  low global luminosities and their characteristic low surface brightness.   
   A direct probe of the evolutionary state of LSB galaxies is the metal abundance 
   in the interstellar medium (ISM). 
   A low abundance generally indicates only limited enrichment
    of the ISM and therefore (in a closed system) a small amount of  evolution.
  
  The high-quality optical spectra from the SDSS permit us to 
estimate the oxygen and nitrogen abundances from electron temperatures
for these two objects.
 Although the host galaxy of SDSS J121811.0+465501.2 has been caught
by the Palomar 48-inch Schmidt survey, and was given
as name  ``MAPS-NGP O\_171\_0165792" in 
the Space Telescope Science Institute (ST ScI) Digitized Schmidt 
Survey (DSS, the reference as 1994DSS...1...0000 in NED), 
there are no any further study for this object. 
Besides the metallicity property, 
by taking advantage of the the $g$- and $r$-band photometric images from SDSS,
we can also study its surface brightness
profiles and obtain its $B$-band central surface brightness etc. 
Then, we identify this galaxy as a low surface brightness edge-on disc galaxy
with  $\mu_0(B)$ $\sim$23.68 mag arcsec$^{-2}$, about 2\,dex lower than 
the Freeman limit value (see Sect.\ref{sec5}).
For the irregular LSB galaxy UGC\,8837, although some studies have been done,
such as optical velocity field and rotation curve 
(Kuzio de Naray et al. 2006; de Blok \& Bosma 2002),
 H$\alpha$ narrow-band observation (James et al. 2004), and 
H\,I observation (Swaters et al. 2002) etc., 
there are no high-quality optical spectroscopic observation and metallicity estimate
for that H~{\sc ii} region coordinated as SDSS J135440.5+535309.6 in this galaxy.
We can estimate the oxygen abundance of this H~{\sc ii} region by using its
SDSS optical spectrum.
  
    This paper is organized as follows. 
    In Sect.2, we show the optical spectra of the two H~{\sc ii} regions
    and the images of their hosts from SDSS,
    and give their coordinates and redshifts.
    In Sect.3, we present the flux measurements of the emission lines,
    and the resulted diagnostic diagrams.
    In Sect.4, we calculate their oxygen and nitrogen abundances 
    using the electron temperature method. 
    In Sect.5, we use the GIM2D software to analyze 
    the $g$- and $r$-band images of the host galaxy of SDSS J121811.0+465501.2,
    and study its morphology and surface brightness.
    The discussions are given in Sect.\ 6.
    We conclude this paper in  Sect.\ 7. 
    Throughout the paper (hereafter), we omit the ``SDSS" before 
    the IAU designations ``J121811.0+465501.2" 
    and ``J135440.5+535309.6" when referring the objects. Since 
    J121811.0+465501.2 is the only H~{\sc ii} region was taken 
    spectral observation by the
    SDSS in its host, and just locates in the center region of the host,
    we also call its host galaxy, the new LSB galaxy, as J121811.0+465501.2
    in the rest parts of the paper.
      A cosmological model
    with $H_0$=70 km s$^{-1}$ Mpc$^{-1}$, $\Omega_M$=0.3 
    and $\Omega_\lambda$=0.7 has been adopted. 
 
\section{The optical spectra and images}
  
  These two extragalactic H~{\sc ii} regions were classified
   as stars in the SDSS-DR4 database, and were found by us 
   in automatic recognition and classification on stellar spectra
   on the basis of wavelet transformation (Liu 2006).
About 243 special stellar spectra with emission lines (H$\alpha$, H$\beta$, H$\gamma$) 
were selected from the SDSS-DR4 database in that study. 
These two objects are among this list, 
and others could be stars with emission-lines mostly or 
unknown objects etc.
The careful studies on other unknown objects are ongoing.
The possible reason to classify them as stars 
could be that they are  
quite bright like point sources with the hosts too faint
in luminosities and surface brightness, 
and they are quite nearby with small redshifts.

\begin{figure}[b]
\centering
\input epsf
\epsfverbosetrue
\epsfxsize 6.5cm
\epsfbox{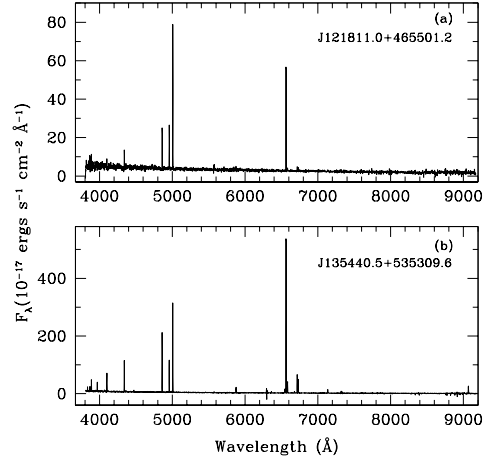}
\caption {The SDSS optical spectra of the two H~{\sc ii} regions:
{\bf (a).} J121811.0 +465501.2 located in the new LSB galaxy;
{\bf (b).} J135440.5+535309.6 located 
in the irregular LSB galaxy UGC\,8837.
}
\label{fig1}
\end{figure}

Figures~\ref{fig1}a,b present their SDSS optical spectra at
rest-frame, 
which show their properties of metal-poor star-forming regions:
(1) almost no continuum and stellar absorption characteristics;
(2) very strong [O~{\sc iii}]4959,5007 and Balmer lines (low and high levels);
(3) very weak [N~{\sc ii}]6548, 6583 and [S~{\sc ii}]6717,6731; 
(4) [O~{\sc iii}]/H$\beta$ ratios are very high, while 
[N~{\sc ii}]/H$\alpha$ and [S~{\sc ii}]/H$\alpha$ ratios are very low.
If [O~{\sc ii}]3727 was not shifted out the observational band, it will
provide more information. 

\begin{figure} [b]
\centering
\input epsf
\epsfverbosetrue
\epsfxsize 6.2cm
\epsfbox{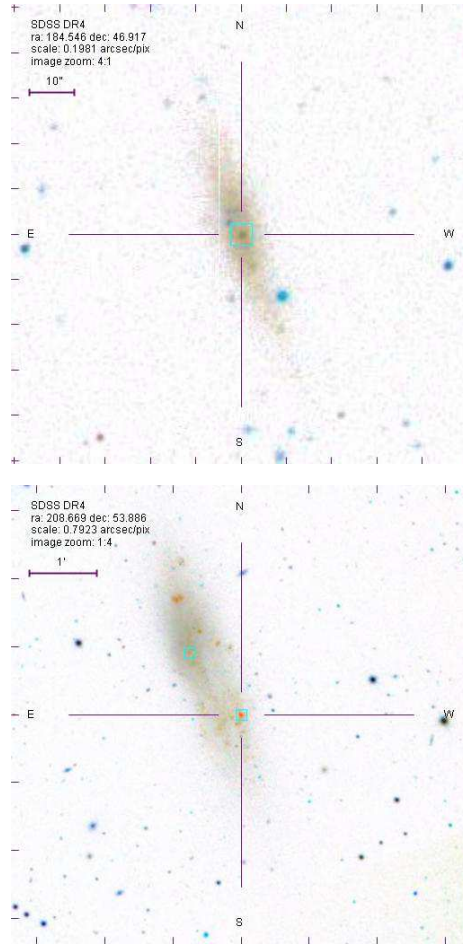}
\caption {The SDSS photometric images of the two objects:
{\bf the upper panel} for the galaxy J121811.0+465501.2 (in size of
1.7$'$$\times$1.7$'$);
{\bf the lower panel} for the irregular LSB galaxy UGC\,8837(in size of
$\sim$6.7$'$$\times$6.7$'$). }
\label{fig2}
\end{figure}

Figures~\ref{fig2}a,b present the images 
taken from the SDSS Finding Chart Tool.
Their RA, DEC (in 2000 epoch) and the image size have been marked on the
left-top corners of the figures. 
The big crosses mark the places where were done 
spectroscopic observations (used in this study) in the galaxies. 
For the galaxy J121811.0+465501.2, the  
spectrum was observed almost on the center part as shown in the upper panel of
Fig.~\ref{fig2}.
This is the first time to obtain its high-quality optical spectrum of this galaxy
since it was named as
``MAPS-NGO 0\_171\_0165792" in the DSS survey. 
The detailed photometric study for this galaxy will be given in Sect.~\ref{sec5}.
The large galaxy presented in the lower panel of Fig.~\ref{fig2}
is the irregular LBG galaxy UGC\,8837, one of its bright H~{\sc ii} regions 
coordinated as J135440.5+535309.6 has also been done spectroscopic observation by the SDSS, 
which could be 
the first high-quality optical spectral observation for this H~{\sc ii} region.
There is another  H~{\sc ii} region coordinated as J135445.63+535403.6 (MJD-PID-FID:
52797-1323-207) in UGC\,8837 has been done spectroscopic observation by the SDSS, but
the S/N ratio of the spectrum is low, which limits us to obtain interesting results 
from it.

The coordinates and redshifts of the two objects are given in Table~1.

\section{Flux measurements and diagnostic diagrams}   

We use SPLOT in IRAF\footnote{IRAF is distributed by the National Optical Astronomical
Observatories, which is operated by the Association of Universities for Research
in Astronomy, Inc., under cooperative agreement with the National Science
Foundation.} package to measure the fluxes of the emission-lines.
The measurements without correction for redding are given in Table~\ref{tab2}.

The dust extinction is estimated from H$\alpha $/H$\beta $:
assuming case B
recombination, with a density of 100\thinspace cm$^{-3}$, a temperature
of 10$^{4}$\thinspace K, and the predicted intrinsic 
H$\alpha $/H$\beta $ ratio of 2.86 (Osterbrock 1989). 
The derived $A_V$ (=$cR_V/1.47$, Seanton 1979) 
of the two objects are given in Table~\ref{tab2}.
As well as some line ratios after correcting the 
fluxes by dust extinction.

\begin{figure}
\centering
\input epsf
\epsfverbosetrue
\epsfxsize 6.5cm
\epsfbox{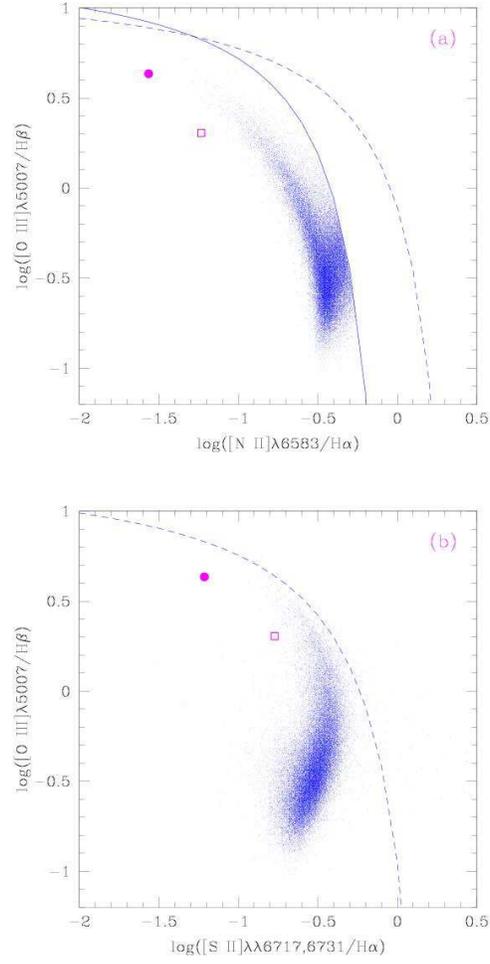}
\caption {The diagnostic diagrams for the two objects. 
{\bf the large solid circle} refers to the object J121811.0 +465501.2, and
{\bf the square} refers to the object J135440.5 +535309.6.
The small points are the star-forming galaxies from the SDSS-DR2.
The lines diagnose the H~{\sc ii} regions and AGNs 
with the solid from MPA/JHU group, and
the dashed from Kewley et al. (2001).
}
\label{figdia}
\end{figure}

Figures~\ref{figdia}a,b present the 
diagnostic diagrams for the two objects 
(the large solid circle for J121811.0+465501.2,
and the square for J135440.5+535309.6) with
the ratios of the extinction-corrected fluxes
[N~{\sc ii}]/H$\alpha$  versus [O~{\sc iii}]/H$\beta$
and [S~{\sc ii}]/H$\alpha$  versus [O~{\sc iii}]/H$\beta$.
The small points in the figures are the star-forming galaxies from
the SDSS (DR2; Tremonti et al. 2004; Liang et al. 2006). 
The lines in the figures
diagnose the H~{\sc ii} regions and AGNs 
with the solid from MPA/JHU\footnote{http://www.mpa-garching.mpg.de/SDSS/} group 
(Kauffmann et al. 2003b), and
the dashed from Kewley et al. (2001).
These diagrams show
the properties  of these 
two objects as H~{\sc ii} regions with low metallicities.

\section{The ``direct" oxygen and nitrogen abundances derived from electron temperature}

 Accurate abundance
measurements for the ionized gas in galaxies require the
determination of the electron temperature ($T_e$), which is
usually obtained from the ratio of auroral to nebular line
intensities, such as 
[O~{\sc iii}]$\lambda\lambda4959,5007$/[O~{\sc iii}]$\lambda$4363.  
This is generally known as the ``direct $T_e$-method" because $T_e$  is
directly inferred from observed line ratios, and sensitive to metallicity. 
  
We adopt the two-zone model for the temperature structure 
within the galactic gas, i.e. 
  $T$([O~{\sc iii}]) (in K, as $t_3$ in 10$^{4}$K) 
is taken to represent the temperature for high-ionization species such as
$O^{++}$, and $T$([O~{\sc ii}]) (in K, as $t_2$ in 10$^{4}$K) 
is used for low-ionization species such as $O^{+}$ (as well as  $N^{+}$).  
Then the total oxygen abundances are derived from the equation:
\begin{equation}
\frac{O}{H} = \frac{O^{++}}{H^+} + \frac{O^{+}}{H^+}.
\end{equation}
 
 From the SDSS spectra of the objects,
  we can obtain their ratios of the auroral ($\lambda$4363)
  to nebular ($\lambda$$\lambda$4959,5007) [O~{\sc iii}]
  lines to estimate their $T$([O~{\sc iii}]) (as $t_3$) temperature 
  in high-ionization regions,
  then
we use the set of equations published by Izotov et al. (2006)
to determinate the ``direct" oxygen abundances in H~{\sc ii} regions for a five-level
atom. 
They used the atomic data from the references listed
in Stasi\'{n}ska (2005).  According to those authors, the electron
temperature $t_3$ and ionic abundances
$O^{++}/H^+$ are estimated as follows:
\begin{equation}
t_3 = \frac{1.432}{{\rm log}((\lambda4959 + \lambda5007)/\lambda4363) - 
{\rm log}C_T},
\end{equation}
where
\begin{equation}
C_T = (8.44 - 1.09t_3 + 0.5t_3^2 - 0.08t_3^3)\frac{1 + 0.0004x_3}{1+0.044x_3},
\label{eqCt}
\end{equation}
with $x_3 = 10^{-4}n_et_3^{-1/2}$, and $n_e$ the electron density in cm$^{-3}$.
Notice that $x_3$ has a very small impact since
it is generally less than 0.1 with $n_e$ $<$ 10$^3$ cm$^{-3}$. 
And
\begin{eqnarray}
12& + &{\rm log}(O^{++}/H^+) = 
{\rm log}(I_{{[O}{III]\lambda4959+\lambda5007}}/I_{H\beta}) + \nonumber \\
  & &  6.200 + \frac{1.251}{t_3} - 0.55{\rm log}t_3 - 0.014t_3.
\end{eqnarray}

The electron densities $n_e$ in the ionized gas of the galaxies are
calculated from the line ratios [S~{\sc ii}]$\lambda6717$/[S~{\sc
  ii}]$\lambda6731$ by using the
  five-level statistical equilibrium
model in the task {\sc TEMDEN}/{\sc STSDAS}
package at $T_e$=10,000K.  
For the object J121811.0+465501.2, we assume the $n_e$ as 10$^2$ cm$^{-3}$,
the typical case of H~{\sc ii}  regions,
since its [S~{\sc ii}] line ratio is outside the physical limit of $\sim$1.431. 

Since [O~{\sc ii}]$\lambda3727$ was shifted out of the observational band,
the ionic abundance $O^+$/$H^+$ cannot be obtained
from direct measurements. 
However, it is reasonable to suppose that oxygen abundance is the same
throughout the H~{\sc ii}  region, the fraction of that oxygen  ionized 
singly or doubly depends on 
the physical conditions in the gas.  
The ratio of $O^+$ to $O^{++}$ (and in principle,
neutral O) within the $H^+$ zone varies with the hardness of the ionizing
spectrum, the gas density, and the O/H abundance itself. 
Since there is 
no observed [O~{\sc ii}]$\lambda3727$ to tell us what be measured, 
some guestimate is
warranted. Therefore, here we assume
$\frac{O^{+}}{H^+}=\frac{O^{++}}{H^+}$,
and then to obtain the total O/H abundance by doubling the
$O^{++}$/$H^+$ abundance. 
It is not particularly reasonable, but
at the accuracy required for this work, this assumption is
tolerable.  
Then the accuracy on the total O/H will be entirely dominated by the $O^+$/$H^+$ 
assumption.  Since this is a factor of two, a more reasonable uncertainty to
quote on the total log(O/H) would be about 0.30 dex.  
Indeed, these H~{\sc ii}  regions are likely to be high ionization 
(typical of low O/H regions), with
low $O^+$/$H^+$ such that the assumed doubling is likely an over-estimate.  
But this assumption won't change the magnitude order of the oxygen abundance.

The $N^+/H^+$ abundances of the objects are estimated by 
using the equation below from 
[N~{\sc ii}]$\lambda$$\lambda$6548,6583 over H$\beta$ ratios
and the $t_2$ temperature (Izotov et al. 2006):
\begin{eqnarray}
&12& + {\rm log}(N^{+}/H^+) = 
{\rm log}(I_{{[N}{II]\lambda6548+\lambda6583}}/I_{H\beta}) + 6.234  \nonumber \\
  & &              + \frac{0.950}{t_2} - 0.42{\rm log}t_2 - 0.027t_2+{\rm log}(1+0.116x_2),
\end{eqnarray}
with $x_2 = 10^{-4}n_et_2^{-1/2}$, and $n_e$ as above.
$t_2$ is the electron temperature in low-ionization, which can be estimated 
from $t_3$ by following an
equation derived by fitting H {\sc ii} region models taken from 
Garnett (1992):   
\begin{equation}
t_2 = 0.7t_3 + 0.3,
\end{equation}
 which is valid over the range 2000K$<$ $T_e$([O~{\sc iii}])$<$ 18,000K,
 and has been used widely. 

  Finally, the oxygen abundances
 are obtained as 12+log(O/H) = 7.88$\pm$0.30 and 7.70$\pm$0.30
 for the objects J121811.0+465501.2
and J135440.5+535309.6, respectively,
which are almost 1/6 of the solar oxygen abundances
when adopting 12+log(O/H)$_{\odot}$$\sim$ 8.66 (Asplund et al. 2005).
Their log($N/H$) abundances are 5.93$\pm$0.05 and 6.38$\pm$0.03, 
 respectively, by assuming $N/H$=$N^+/H^+$. 
 Then the derived log(N/O) abundances are 
 -1.65$\pm$0.30 and -1.01$\pm$0.30, respectively. The abundances and
some related line ratios, $n_e$, $t_3$, $t_2$ 
are also given in Table~\ref{tab2}. The error-bars 
come from the uncertainties of the measurements, without considering
the uncertainties of calibrations and conversions.

\section{Image analysis of J121811.0+465501.2: a new LSB galaxy}
\label{sec5}

We use the publicly available GIM2D package (Simard et al. 2002)
to find best-fit PSF-convolved, two-dimensional (2-D) modelling of
the target galaxy J121811.0+465501.2. The 2-D fitting deals with real
galaxy images directly, and develops a projected model to
approach real light distribution of galaxies, which can provide
extra bulge/disc separation. Performing light fitting in 2-D
allows proper weighting of data points reflecting uncertainties
due to photon statistics, flat fielding uncertainties, sky level
uncertainties and model imperfection (de Jong et al. 2004).

In this analysis, we use the SDSS $g-$ and $r-$band images 
since the images in $u-$ and $z-$band are relatively shallow
whereas $i-$band image may suffer from the `red halo' effect
(Michard 2002; Wu et al. 2005). Since the main source of
errors in determining surface brightness profiles for this class
of galaxies is the uncertainty in sky estimation, we have
performed our own sky subtraction on the images carefully. A detailed
description of our method has been provided by Liu et al. (2007).

We found the light of galaxy J121811.0+465501.2 can be modeled
well by a single exponential disc profile of the form:

\begin{equation}
\Sigma (r) = \Sigma_0 exp(-{{r}\over{h}}),
\end{equation}
where $\Sigma_0$ is the face-on central surface brightness
in linear units ($M_{\odot}$ pc$^{-2}$), and $h$ is the
scale length along the semi-major axis.
This equation can be converted to logarithmic units as: 
\begin{equation}
\mu (r) = \mu_0 + 1.086({{r}\over{h}}),
\end{equation}
where $\mu_0$ is the face-on central surface brightness of 
the disc in logarithmic units
(mag arcsec$^{-2}$) (de Blok et al. 1995). 

\begin{figure*}
   \centering
   \includegraphics[width=12cm]{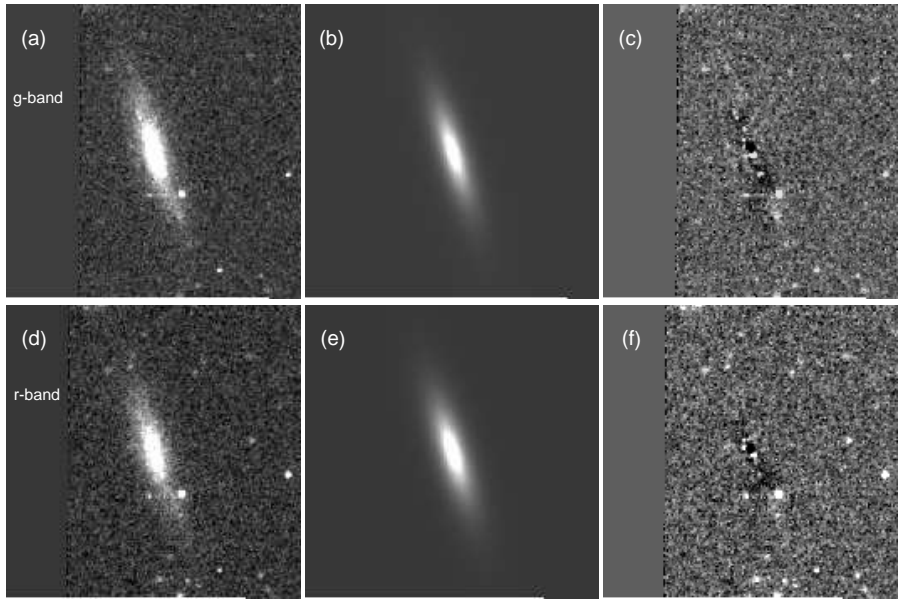}
      \caption{ The postage $g-$ and $r-$band images with size of 1.7$'$$\times$1.7$'$ 
      of the newly found LSB galaxy SDSS J121811.0+465501.2:
      {\bf (a).} the $g$-band postage image;
      {\bf (b).} the best-fit model image for the $g$-band image;
      {\bf (c).} the residual image in the $g$-band; 
      {\bf (d).} the $r$-band postage image;
      {\bf (e).} the best-fit model image for the $r$-band image;
      {\bf (f).}  the residual image in the $r$-band.
              }
         \label{fig4}
   \end{figure*}

The analysis results are given in Figs.~\ref{fig4}a-f, including
the trimmed postage images, the best-fit model images, and the
residual images, all in both of the $g$ and $r$ bands. The
reduced $\chi^2$ for the model fittings
are 1.071 and 1.075 in $g-$ and $r-$bands,
respectively. The resulted parameters have been given in 
Table~\ref{tab3}, including the central surface brightness
$\mu_0$, the scale-length $h$ of the disc, 
the inclination angle etc., all in the two bands. 
By using the conversion of Smith et al. (2002), we can
estimate its $\mu_0(B)$ = 23.68 and $\mu_0(V)$ = 22.96 mag
arcsec$^{-2}$. These results show the galaxy J121811.0+465501.2 is
a nearly edge-on LSB galaxy, with an exponential disc
component and almost without bulge. 

In the analysis, this disk model is assumed
to be infinitely thin. The total flux in this model is given by:
\begin{equation}
F_{tot} = 2 {\pi} h^2 \Sigma_0.
\end{equation}
Then its total apparent magnitude can be determined from this equation and converting
to a logarithmic scale. The obtained apparent magnitudes 
in the $g$ and $r$ bands are given in Table~\ref{tab3}, as well as
the $B$- and $V$-band magnitudes calculated by using the conversion 
of Smith et al. (2002). 
To estimate the absolute magnitude, we calculate the 
dust extinction inside the galaxies in $g$ and $r$ bands
from $A_V$ (see Table~\ref{tab2}), and
also consider the Galactic foreground
extinction following the reddening maps of Schlegel et al. (1998).
The redshift has been corrected for the motion of the Sun and Local
Group using the methods of Courteau \& van den Bergh (1999)
and Blanton et al. (2005), and given as $z_{corr}$ in 
Table~\ref{tab3}. 
One point should be noticed that we neglect the correction for 
the peculiar velocity of the galaxy, which could be estimated by maximizing the likelihood
density as Blanton et al. (2005). 
They commented that the typical corrections of this
are 200-300 km s$^{-1}$ with dispersion of 150 km s$^{-1}$.
We then consider 300 km s$^{-1}$ as the velocity uncertainty for estimating luminosity distances
and magnitudes.
Therefore, the
absolute $B-$magnitude, $M_B$, can be calculated as $-$13.51 mag
from the apparent $B-$magnitude after correction for the
extinction and considering the luminosity distance.
As the redshift of this galaxy is low, no $k-$correction is applied here.
 
 This galaxy is similar in size scale and distance to the object
 $F564-V3$ studied by de Blok et al. (1995).
They found $F564-V3$ has a scale length of only 0.3\,kpc, 
total magnitude $M_T=-$11.77 and luminosity distance $D_L$=6Mpc (with $H_0$=100 km
s$^{-1}$ Mpc$^{-1}$),
 which is the 
smallest and faintest galaxy in their sample, and could be a true dwarf galaxy. 
Similarly, our LSB galaxy J121811.0+465501.2 has a scale length of $\sim$0.4\,kpc, 
$M_B=-13.51$ and $D_L\sim$9.16Mpc (with $H_0$=70 km s$^{-1}$ Mpc$^{-1}$). 
The difference between these two objects is that $F564-V3$ is amorphous and diffuse, 
and its surface brightness profile could not be matched by a perfectly exponential disc,
while our galaxy J121811.0+465501.2 does show a perfectly exponential disc (almost without bulge).
These analyses show that this new LSB galaxy 
could be a true nearby dwarf galaxy with an nearly exponential disc shown as edge-on to us.

\section{Discussions}

  It is difficult to detect LSB galaxies owing to their intrinsically low 
  global luminosities and low surface brightness.  Several samples of LSB objects, identified
  with different criteria and on either photographic plates,
  or on CCD images, have been published during last 20 years (for a review see Impey \& Bothun
  1997; Dalcanton et al. 1997). 
  Recently,  Kniazev et al. (2004) developed a method to use imaging data from the SDSS to search
  for LSB galaxies, and tested their method by using 92 galaxies from the catalog 
  of Impey et al. (1996). 
  Haberzettl et al. (2007) also reported on the photometric results of a search for LSB galaxies in
  a 0.76 deg$^2$ field centered on the Hubble Deep Field-South.
  In this study, we present a new LSB galaxy found from the SDSS-DR4 occasionally.

  Metallicity is an efficient indicator to probe the evolutionary status and 
  history of LSB galaxy. 
     The first measurements of the oxygen abundances in H~{\sc ii} regions of LSB galaxies were
    presented in McGaugh (1994).           
       They found that, in general, LSB galaxies are metal-poor ($Z<1/3Z_{\odot}$),
       which indicates that LSB galaxies evolve slowly, forming relatively few 
       stars over a Hubble time. 
     	Burkholder et al. (2001) 
	also found that the average oxygen abundances
	of their 17 LSB galaxies is one-fifth solar.
	de Blok \& Hulst (1998) found that 
	most of the 64 H~{\sc ii} regions in 12 LSB galaxies
    have oxygen abundances that are $\sim$0.5 to 0.1 solar. 
    The two extragalactic H~{\sc ii} regions in this study 
    have about 6 times lower oxygen abundances
    than the solar.
    Moreover, no strong radial oxygen abundance
    gradients are found in the LSB galaxies, which is
    supporting the picture of their quiescent and sporadic evolution, which is 
	different from the HSB galaxies with the steeper gradients 
	(Vila-Costas \& Edmunds 1992; Zaritsky, Kennicutt, Huchra 1994; 
	Henry \& Howard 1995; Kennicutt \& Garnett 1996). 	   

    About the stellar populations of LSBGs,
       Galaz et al. (2005) computed the Lick/IDS indices for the light emitted
     by the bulges hosted by the 19 face-on LSB galaxies, compared with 
     a sample of 14 face-on HSB galaxies. Their results
     indicate that bulges of LSBGs are metal poorer, compared to those hosted by
     HSBGs, supporting the hypothesis that stellar formation rates have been
     stable low during the lifetime  of these galaxies. 
    Bell et al. (1999) used the NIR K$'$ and optical images of a sample of 
    five red and blue LSB disc galaxies
   to study the stellar populations of them. They found that the blue LSBGs 
   are well described by
   models with low, roughly constant star formation rates, 
   whereas red LSBGs are better described by a `faded disc' scenario.
      
   de Blok et al. (1995) studied the surface
brightness profiles and colors of 21 late-type LSBGs.
 Combination of their
data with those of McGaugh (1992) yields median values of 
respectively $B$-band central surface brightness $\mu_0(B)$=23.4 mag
arcsec$^{-2}$ and scale length $h=3.2$ kpc. The median of the surface brightness is thus
shifted more than 6$\sigma$ from the Freeman (1970) value 
$\mu_0(B)$=21.65 mag arcsec$^{-2}$.
The favored interpretation of blue colors in LSB galaxies is that of
low metallicity coupled with sporadic star formation. 
Moreover, the blue colors show that the LSB galaxies are not faded discs that have 
no current star formation.
For our galaxy J121811.0+465501.2,  
we have tried to do the same one dimension fit for its surface brightness profile
by using the ISOPHOTE/ELLIPSE task in IRAF package.
It also confirm well its exponential disc component of a LSBG without bulge.
But as the inclination angle is large ($>$70 degree), the integrated light 
at a given effective radius could be a projected
result from slightly different radii. Therefore, we prefer to present the
2D-fittings by using GIM2D on its photometry images, 
which have been given in Sect.~\ref{sec5}.

\section{Conclusions}
\label{sec7}  

  We present a new Low Surface Brightness Galaxy SDSS J121811.0+465501.2
 serendipitously found from the SDSS-DR4 database, and
 one of its H~{\sc ii} region was classified as a star by the spectral
 classification of SDSS.  
 This galaxy has $B$-band central surface brightness as
 $\mu_0(B)$$\sim$ 23.68 mag arcsec$^{-2}$, which is $\sim$2\,dex fainter than 
 the Freeman limit of classifying as LSBGs, and
 confirms that it is a LSB galaxy. It has
 an exponential disc and almost no bulge. 
 And it could be a true nearby dwarf galaxy
 with a scale length of 0.4\,kpc and $M_B$$\sim -$13.51, $D_L\sim$9.16Mpc 
 (with $H_0$=70 km s$^{-1}$ Mpc$^{-1}$).
 Its central region shows obvious optical emission lines,
 which indicates that there exist star-forming activities presently.
 We obtain its oxygen abundance as 12+log(O/H)$\sim$7.88
 from electron temperature on the basis of the SDSS optical spectrum. 
 We also estimate the oxygen abundance of another 
 H~{\sc ii} region J135440.5+535309.6 located in the irregular
LSB galaxy UGC\,8837, which was found from the 
same procedure from SDSS-DR4 occasionally
as for the object J121811.0+465501.2. 
It has 12+log(O/H)$\sim$7.70, similar to J121811.0+465501.2. 
Their oxygen abundances are about six times more metal-poor than 
the solar by assuming 12+log(O/H)$_{\odot}$$\sim$8.66. 
These low metallicities confirm other
studies for H~{\sc ii} regions of LSBGs, 
in which they showed that the LSBGs have low metallicities.
 
In summary, in the edge-on disc LSB galaxy and the irregular LSB galaxy UGC\,8837, 
there are massive stars forming presently, but the metallicities there are very low.
These could support the picture of quiescent and sporadic evolution,
and the less efficient metal-enrichment of the LSBGs.

 These two extragalactic objects are among the list of 243 ``stars" with special spectra, 
which were selected from the SDSS-DR4 database in 
automatic recognition and classification on stellar spectra
 to search for emission-line stars. 
 Most are identified as emission-line stars, such as CV stars and others,
 but left some interesting objects as we studied here.
 We will check the SDSS-DR5 to find more such interesting objects.

\acknowledgments
  We specially thank our referee Dr. Stacy McGaugh for his very valuable
  suggestions and comments, which help well in improving this work.
  We thank Profs. Shude Mao, Xiaoyang Xia and Zugan Deng for very interesting
  discussions.
  This work was supported by the Natural Science Foundation of China
 (NSFC) Foundation under No.10403006, 10433010, 10373005, 10573022, 10333060, 
 and 10521001.


 
\clearpage

\begin{table*}
\centering
\caption{\label{tab1} \footnotesize Coordinates and redshifts of the two
objects. The redshift is the average value of the three
calculated from three strong emission lines on the spectrum, 
H$\beta$, [O~{\sc iii}]$\lambda$5007 and H$\alpha$.}
\label{tab1}
\scriptsize
\begin{tabular}{lrrrrrr}
\hline \hline
\noalign{\smallskip}
(1)  & (2) & (3) & (4) & (5) & (6)  & (7)  \\       \hline            
\noalign{\smallskip}
 MJD-PID-FID    &  IAU designation     &   R.A. (2000) & DEC. (2000) & redshift $z$  & $cz$ (km s$^{-1}$)  & host galaxy  \\ \hline
53117-1451-181  &  J121811.0+465501.2  &  12 18 11.04 (184.546)  & +46 55 01.2 (46.917) & 0.00157 & 472.46   & new LSBG            \\ 
52465-1043-613  &  J135440.5+535309.6  &  13 54 40.56 (208.669) & +53 53 09.6 (53.886) & 0.00067 & 210.38   & UGC 8837                  \\ \hline
\end{tabular}
\end{table*}

\begin{table}
\centering
\caption{\label{tab1} \footnotesize The flux measurements of emission lines
(in units of 10$^{-17}$ erg s$^{-1}$ cm$^{-2}$ \AA$^{-1}$) of the two objects,
some related line-ratios corrected by dust extinction,
and the derived electron temperature and abundances are given as well.}
\label{tab2}
\scriptsize
\begin{tabular}{lrr}
\hline \hline
\noalign{\smallskip}
line               & J121811.0+465501.2 & J135440.5+535309.6   \\ \hline
4861  H$\beta$       &  72.51$\pm$0.52  &  670.04$\pm$3.12   \\ 
4959  [O~{\sc iii}]   &  75.20$\pm$1.69  &  347.32$\pm$1.79   \\ 
5007  [O~{\sc iii}]   & 236.78$\pm$1.88  & 1040.76$\pm$1.86   \\ 
4363  [O~{\sc iii}]   &   3.96$\pm$0.38  &   12.82$\pm$1.28  \\ 
6563  H$\alpha$      & 239.04$\pm$0.14  & 2300.34$\pm$6.14   \\ 
6583  [N~{\sc ii}]    &   7.26$\pm$0.31  & 163.09 $\pm$1.47   \\ 
6717  [S~{\sc ii}]    &  10.53$\pm$0.01  &  272.46$\pm$0.77   \\ 
6731  [S~{\sc ii}]    &   6.46$\pm$0.01  &  199.27$\pm$0.25   \\  \hline
$A_V$                          &  0.35$\pm$0.02   &  0.45$\pm$0.02 \\ 
log([N~{\sc ii}]$\lambda$6583/H$\alpha$)   & -1.52$\pm$0.02   & -1.16$\pm$0.01  \\ 
log([S~{\sc ii}]6717,6731/H$\alpha$)   & -1.15$\pm$0.01   & -0.69$\pm$0.01  \\ 
log([O~{\sc iii}]4959,5007/H$\beta$)   &  0.63$\pm$0.01   &  0.30$\pm$0.01 \\  \hline
log(${[O~{\sc III}]4959,5007}\over [O~{\sc III}]4363$)  & 1.87$\pm$0.04  &   2.00$\pm$0.04 \\ 
{[S~{\sc II}]$\lambda$6717/[S~{\sc II}]$\lambda$6731}   & 1.63$\pm$0.01  & 1.37$\pm$0.01 \\ 
$n_e$ (cm$^{-3}$)    & ---  & 55.33$\pm$1.81  \\ 
$t_3$ (10$^{4}$K)    &  1.4487$\pm$0.0536    &  1.2872$\pm$0.0500  \\ 
$t_2$ (10$^{4}$K)    &  1.3141$\pm$0.0375    &  1.2011$\pm$0.0350  \\ 
 12+ log(${O^{++}}/{H^+}$)   & 7.58$\pm$0.05   & 7.40$\pm$0.05  \\ 
  12+log(${O}/{H}$)          & 7.88$\pm$0.30   & 7.70$\pm$0.30   \\ 
  12+log(${N}/{H}$)          & 5.93$\pm$0.05   & 6.38$\pm$0.03   \\ 
  log(N/O)                  & -1.65$\pm$0.30  & -1.01$\pm$0.30  \\   \hline
 \end{tabular}
\end{table}

\begin{table}
\centering \caption{\label{tab1} \footnotesize 
The parameters of the galaxy J121811.0+465501.2
about its morphology, surface brightness and magnitudes etc.
by performing GIM2D on its SDSS $g$- and $r$-band images. }
\label{tab3} \scriptsize
\begin{tabular}{lrr}
\hline \hline
\noalign{\smallskip}
    \multicolumn{3}{c}    {J121811.0+465501.2}        \\ \hline
                   &   $g$-band     &   $r$-band   \\ \hline
$\mu_0$ (mag arcsec$^{-2}$)    &  23.39$\pm$0.05  & 23.13$\pm$0.07 \\
$h$  (arcsec)              &  9.12$\pm$0.14    &  8.98$\pm$0.18  \\
$h'$  (kpc)                 &  0.40$\pm$0.18    &  0.40$\pm$0.18    \\
inclination angle (deg)     & 75.5$\pm$0.3    & 74.0$\pm$0.4 \\
$m_T$ (total apparent mag)   & 16.59$\pm$0.01  & 16.36$\pm$0.02  \\   
$M_T$ (total absolute mag)   & -13.72$\pm$0.84    & -13.81$\pm$0.84 \\   \hline
$\mu_0(B)$ (mag arcsec$^{-2}$)  &  \multicolumn{2}{c} {23.68$\pm$0.08}   \\
$\mu_0(V)$ (mag arcsec$^{-2}$)  & \multicolumn{2}{c} {22.96$\pm$0.11}    \\
$B_T$ (total apparent B mag) & \multicolumn{2}{c} {16.87$\pm$0.02}   \\  
$V_T$ (total apparent V mag) & \multicolumn{2}{c} {16.43$\pm$0.03}   \\  
$z_{corr}$ (corrected redshift)  & \multicolumn{2}{c} {0.002135$\pm$0.001000}  \\ 
$D_L$ (luminosity distance in Mpc)  & \multicolumn{2}{c} {9.1647$\pm$4.3023}  \\  
$D_A$ (angle distance in Mpc)  & \multicolumn{2}{c} {9.1257$\pm$4.2573}  \\ 
$M_B$ (total absolute B mag)   & \multicolumn{2}{c} {-13.51$\pm$0.84}   \\
$M_V$ (total absolute V mag)   & \multicolumn{2}{c} {-13.80$\pm$0.84}   \\ \hline
 \end{tabular}
\tablenotetext{}{Notes.-The uncertainties of luminosity distance, angle distance, $h'$,
and absolute magnitudes come from the uncertainty of the corrected redshift,
which was estimated as the maximum value of the typical correction 
for the peculiar velocity of
nearby galaxies, about 300 km s$^{-1}$ mentioned in Blanton et al. (2005).}
\end{table}

\end{document}